\documentclass{mem}
\usepackage{natbib}
\usepackage{txfonts}
\usepackage{balance}
\usepackage{graphicx}
\usepackage{flushend}
\usepackage{xcolor}
\definecolor{xlinkcolor}{cmyk}{1,1,0,0}
\usepackage[
  bookmarks=true,         
  pdfnewwindow=true,      
  breaklinks=true,
  colorlinks=true,        
  linkcolor=xlinkcolor,   
  citecolor=xlinkcolor,   
  filecolor=xlinkcolor,   
  urlcolor=xlinkcolor,    
  final=true,
]{hyperref}
\idline{75}{282}
\begin{document}
\def\teff{$T\rm_{eff }$}
\def\kms{$\mathrm {km s}^{-1}$}

\title{
Lights in the Dark: Globular clusters as dark matter tracers
}

   \subtitle{}

\author{
Lucas\ M.\ Valenzuela\inst{1} 
          }

\institute{
Universitäts-Sternwarte, Fakultät für Physik, Ludwig-Maximilians-Universität München, Scheinerstr. 1, 81679 München, Germany
\\ \email{lval@usm.lmu.de}
}

\authorrunning{Valenzuela}

\titlerunning{GCs as DM tracers}

\date{Received: Day Month Year; Accepted: Day Month Year}

\abstract{
A long-standing observed curiosity of globular clusters (GCs) has been that both the number and total mass of GCs in a galaxy are linearly correlated with the galaxy’s virial mass, whereas its stellar component shows no such linear correlation. This work expands on an empirical model for the numbers and ages of GCs in galaxies presented by \citet{valenzuela+21} that is consistent with recent observational data from massive elliptical galaxies down to the dwarf galaxy regime. Applying the model to simulations, GC numbers are shown to be excellent tracers for the dark matter (DM) virial mass, even when distinct formation mechanisms are employed for blue and red GCs. Furthermore, the amount of DM smooth accretion is encoded in the GC abundances, therefore providing a measure for an otherwise nearly untraceable component of the formation history of galaxies.
}
\maketitle{}

\section{Introduction}

As some of the oldest objects found in the Universe, globular clusters (GCs) have long been studied in the Milky Way and many other galaxies due to being dense clusters of stars and therefore having large luminosities. Their old age and the fact that they are found in galaxies of all types and masses enables them to be used as tracers for galaxy formation, evolution, and multiple present-day properties. For instance, it has been found from observations that the number and total mass of GCs correlate linearly with the virial mass of the host system \citep[e.g.,][]{blakeslee+97,harris&blakeslee&harris+17,forbes+18}.

Despite GCs having been studied in detail in galaxies, their formation is still poorly understood. Observations have shown that GCs come in two groups of colors and metallicities: red, metal rich GCs, and blue, metal poor GCs \citep[e.g.,][]{usher+12}. Both types are found in most massive galaxies, where the red GCs tend to populate the inner regions and trace the stellar light distribution \citep[e.g,][]{peng+04,schuberth+10,dolfi+21}, whereas blue GCs lie further out and can have a different spatial distribution \citep[e.g.,][]{forbes+97,schuberth+10}. These differences led to the idea that red and blue GCs are formed through different pathways: blue GCs may tend to form in small galaxies at early times, later being accreted onto more massive galaxies \citep[e.g.,][]{cote+98,schuberth+10,forbes&remus18}, whereas red GCs may later form after the gas has been enriched, for example in starbursts caused by major wet mergers \citep[e.g.,][]{ashman&zepf92,goudfrooij+01,schuberth+10,harris&ciccone+17}.

Empirical models have proven to be useful to study GC populations and their formation in a statistical manner. These models are built on merger trees obtained from cosmological simulations or through generating mock merger trees. The models have generally either assumed GCs to form at very early times \citep[e.g.,][]{beasley+02,boylan_kolchin17,burkert&forbes20} or to form by high accretion rates with sufficient available gas \citep[e.g.,][]{beasley+02,choksi+18,el_badry+19}. These studies have shown the linear relation between the GC abundances and the halo virial masses to result from hierarchical merging and the central limit theorem. However, the former models do not take into consideration the younger GC populations and the latter models only find GC populations in the more massive galaxies, which does not agree with the results from \citet{forbes+18}.

To address this issue, \citet{valenzuela+21} introduced a model that combines both pathways of GC formation, finding that only the combination of the two can reproduce the observations.
This work expands on their findings by considering the individual contributions to the age distribution of the GC formation pathways and by quantifying the relation between GC abundances and smooth accretion.

\section{Globular Cluster Model}

The two GC formation pathways employed by the model from \citet{valenzuela+21} follow the idea that GCs are either formed in small galaxies at early times or in gas-rich mergers of more massive galaxies. Both pathway models are kept as simple as possible to allow probing the whole parameter space, meaning that the number of free parameters is kept to a minimum. These models track the number of GCs in each galaxy and the individual GC formation times. They were run on the merger tree of a dark matter-only cosmological simulation with a side length of $30\,\mathrm{Mpc}$ within the empirical model \textsc{emerge} \citep{moster+18}. The stellar and gas components are empirically inferred from the dark matter halos and their merger trees by \textsc{emerge}. For more details on this see \citet{moster+18}.

\paragraph{\textbf{\textrm{Small Halo Pathway}}}

\begin{figure*}[t!]
\resizebox{\hsize}{!}{\includegraphics[clip=true]{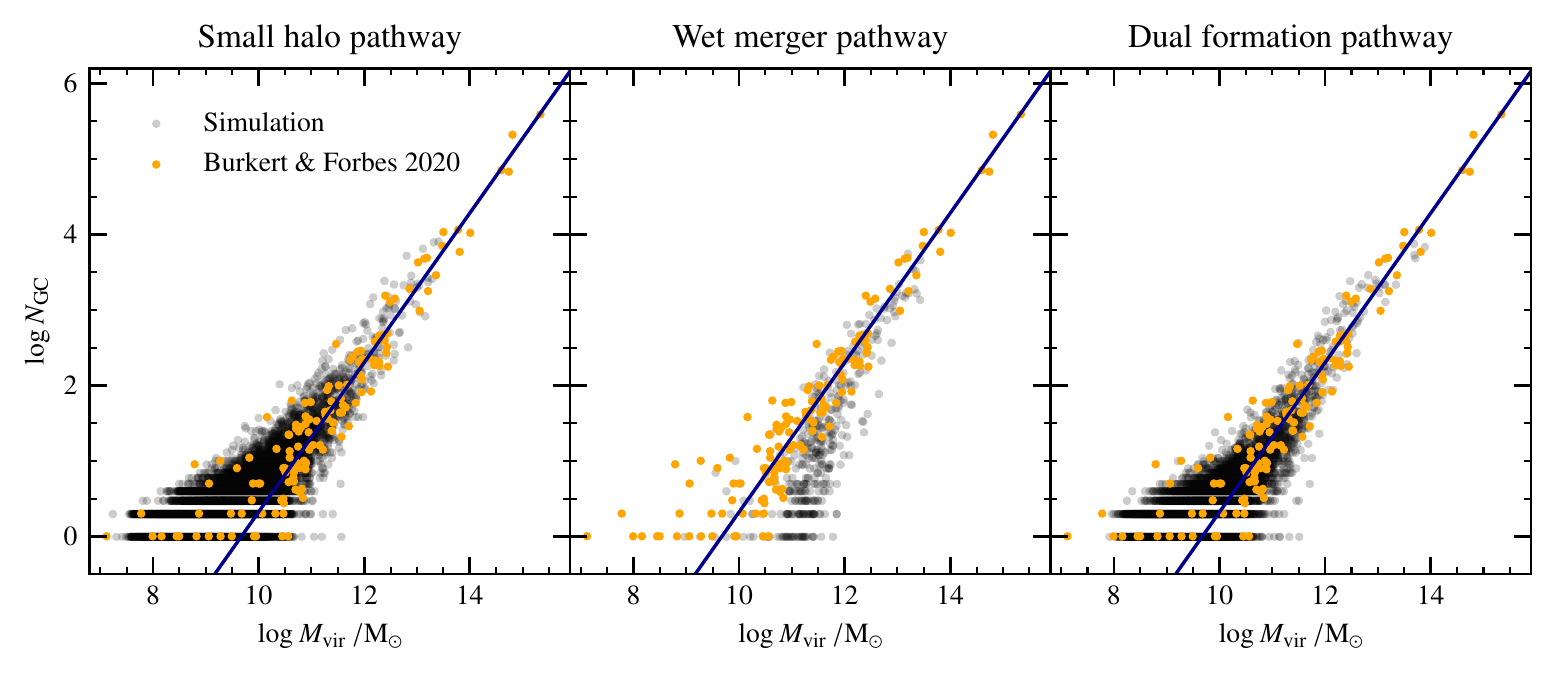}}
\caption{\footnotesize
Relation between the virial mass and the number of GCs of galaxies in observations \citep{burkert&forbes20} and from the GC models. From left to right the first model, the second model, and the combined models, respectively. The solid blue lines indicate the fit relation to the observed galaxies. The virial masses of the simulated galaxies are shown with a random measurement uncertainty of $0.3$, $0.25$, and $0.25\,\mathrm{dex}$, respectively.
}
\label{fig:ngc}
\end{figure*}

\begin{figure*}[t!]
\resizebox{\hsize}{!}{\includegraphics[clip=true]{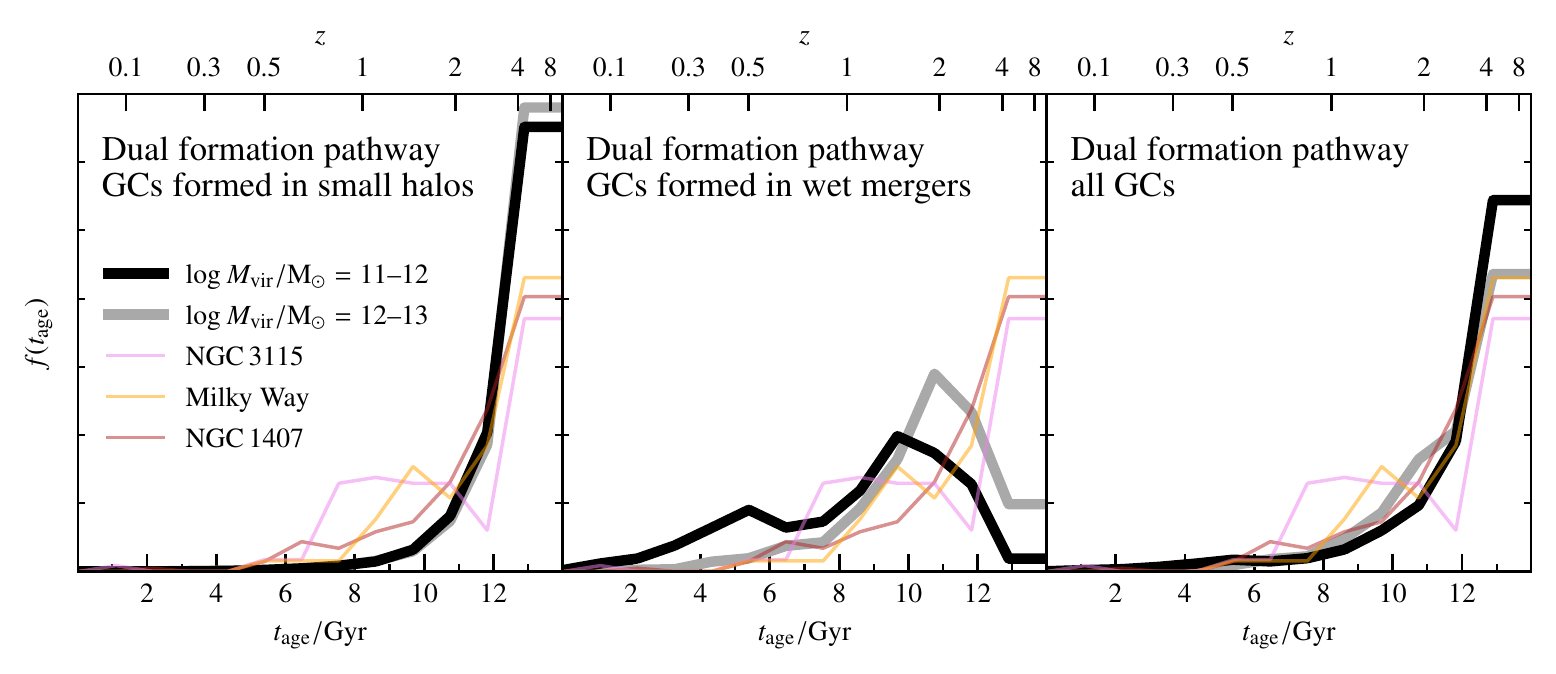}}
\caption{\footnotesize
Age distributions of three local galaxies including the Milky Way \citep{usher+19} compared to the mean age distributions of the modeled GC populations in the two relevant virial mass bins (i.e., the age probability density function of the GCs found in an individual observed galaxy in color, and of all GCs around galaxies within the respective virial mass bins in gray). The GC age $t_\mathrm{age}$ indicates the lookback time of the GC formation, with the corresponding formation redshift $z$ shown on the top x-axes. All panels show the GCs from the dual formation pathway model, where the first two panels show only the GCs formed through the small halo pathway and the wet merger pathway, respectively. The third panel shows the total GC population.
}
\label{fig:ages}
\end{figure*}

In the first model, GCs are formed as soon as the virial mass of a galaxy surpasses a threshold seeding mass, the first free parameter. On average one GC is formed: either 0, 1, or 2 are formed at random.
By adapting the threshold seeding mass, the observed relation between the virial mass and the number of GCs from \citet{burkert&forbes20} can be reproduced (left panel of Fig.~\ref{fig:ngc}). However, when comparing the cumulative age distribution of the GCs, the model lacks the younger GCs found in local galaxies by \citet{usher+19} (left panel of Fig.~\ref{fig:ages}).
Clearly, the model forming GCs in small halos is not sufficient for explaining the observed GC populations.

\paragraph{\textbf{\textrm{Wet Merger Pathway}}}

The second model follows the one introduced by \citet{choksi+18}: when the mass accretion rate surpasses an accretion rate threshold (the first free parameter), GC formation is triggered. The total GC mass is determined from the cold gas mass through a conversion factor (the second free parameter), from which the number of GCs formed is derived by an exponential cluster initial mass function with a slope of $-2$. This model only forms GCs in galaxies with virial masses above roughly $10^{10.5}\,\mathrm{M}_\odot$ and therefore lacks the GCs in low-mass galaxies as seen in the middle panel of Fig.~\ref{fig:ngc}. This model is therefore also not sufficient to explain the GC observations.

\paragraph{\textbf{\textrm{Dual Formation Pathway}}}

Running both models at the same time makes it possible to reproduce the GC number-virial mass relation again (right panel of Fig.~\ref{fig:ngc}), and in contrast to the first model, now also results in a similar age distribution to that of the Milky Way and NGC\,1407 for simulated galaxies of the corresponding virial mass bins (right panel of Fig.~\ref{fig:ages}). This arises from the much younger GCs formed in wet mergers (middle panel of Fig.~\ref{fig:ages}). In summary, the first pathway contributes the older GC population, whereas the second forms younger GCs.
Varying the parameters either leads to an increasingly non-linear relation between the number of GCs and the virial mass, or to much too old GC populations, or to unreasonably low or high thresholds for GC formation in wet mergers (for more details, see \citealp{valenzuela+21}).

\section{Smooth Accretion}

\begin{figure}[t!]
\resizebox{\hsize}{!}{\includegraphics[clip=true]{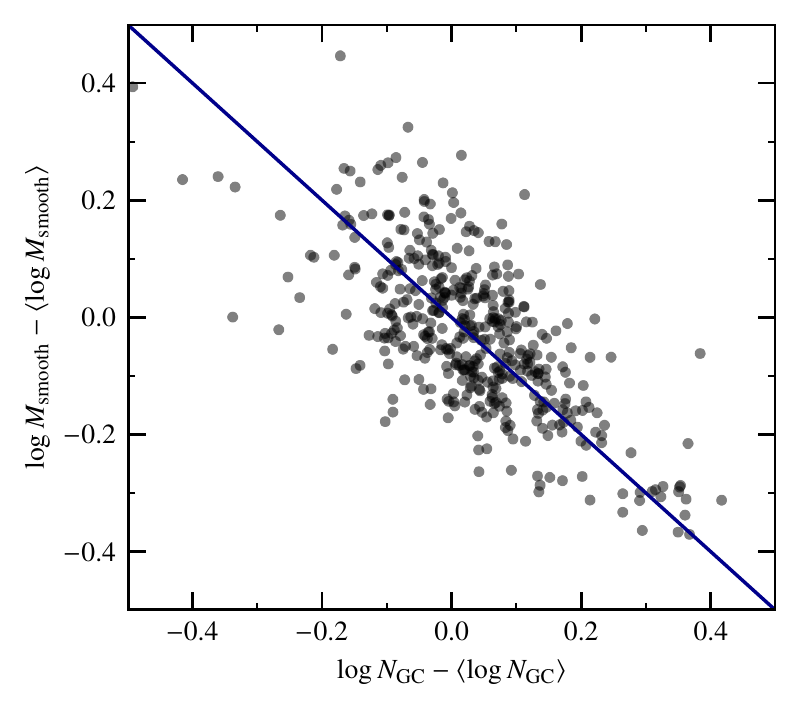}}
\caption{\footnotesize
Relation between the overabundance of GCs and the deviation from the average smoothly accreted matter of the simulated galaxies with virial masses above $10^{11}\,\mathrm{M}_\odot$. The solid blue line indicates the negative 1:1 relation.
}
\label{fig:smooth}
\end{figure}

Despite the limited number of free parameters, the model results in a non-negligible scatter of GCs for a given virial mass. The main driver of this scatter is the amount of smoothly accreted matter throughout a galaxy's formation history, where anything that was accreted with a virial mass below $10^{8.8}\,\mathrm{M}_\odot$ (the GC seeding mass of the first model) is considered to have been smoothly accreted. The relation between the logarithmic over- or underabundance and the above- or below-average smooth accretion for galaxies with the same virial masses is anti-correlated with a slope of roughly $-1$ (Fig.~\ref{fig:smooth}). In conclusion, this means that galaxies with more GCs than average have had less smooth accretion than other galaxies of the same virial mass and vice versa.

\section{Conclusion}

By combining two common approaches to empirically modeling GCs in a dual formation scenario, GC numbers and ages can be found that agree with those observed in galaxies of all masses. The individual models either fail to obtain young GC populations or lack GCs in dwarf galaxies. The necessity of two formation pathways suggests that also the spatial and kinematic properties of GCs may trace different aspects of galaxies. First results on red GCs tracing the stellar light and blue GCs tracing the dark matter halo (Kluge et al., in prep.) are a good indication for distinct formation pathways of these GC populations, as proposed by the model.
Furthermore, the over- or underabundance of GCs with respect to the average number found in galaxies of the same virial mass gives a measure for the smooth accretion onto a galaxy, a nearly untraceable property of galaxy formation history.

\begin{acknowledgements}
LMV acknowledges support by the COMPLEX
project from the European Research Council (ERC) under the European Union’s
Horizon 2020 research and innovation program grant agreement ERC-2019-AdG
882679.
This research was supported by the Excellence Cluster ORIGINS, funded by the
Deutsche Forschungsgemeinschaft under Germany’s Excellence Strategy -- EXC2094-390783311.
\end{acknowledgements}

\bibliographystyle{aa}
\bibliography{bib}

\end{document}